\documentclass[aps,prb,twocolumn,showpacs,superscriptaddress,floatfix,10pt]{revtex4-1}
\usepackage{graphicx}
\usepackage{amsmath}
\usepackage{epsfig}
\usepackage{helvet}
\usepackage{amssymb}
\usepackage{framed}
\usepackage{enumitem}
\usepackage{todonotes}
\usepackage{comment}
\usepackage{float}
\usepackage{bbding}
\usepackage{pifont}
\usepackage{wasysym}
\usepackage[bookmarksnumbered=true,hypertexnames=false,colorlinks=true,linkcolor=blue,urlcolor=blue,citecolor=blue,anchorcolor=green]{hyperref}

\newcommand{\bk}[1]{\mbox{$\langle #1 \rangle$}}

\newcommand{\frw}[1]{$\overset{\lower0.5em\hbox{$\smash{\scriptscriptstyle\smile}$}} #1$}

\begin{document}

\title{Tensor Renormalization Group Centered About a Core Tensor}
\author{Wangwei Lan}
\author{Glen Evenbly}
\affiliation{School of Physics, Georgia Institute of Technology, Atlanta, GA 30332, USA}
\affiliation{D\'epartement de Physique and Institut Quantique, Universit\'e de Sherbrooke, Qu\'ebec J1K 2R1, Canada
}
\email{lan.wangwei@gmail.com}
\date{\today}
\setcitestyle{square}
\begin{abstract}
We propose a modified form of a tensor renormalization group algorithm for evaluating partition functions of classical statistical mechanical models on 2D lattices. This algorithm coarse-grains only the rows and columns of the lattice adjacent to a single core tensor at each step, such that the lattice size shrinks linearly with the number of coarse-graining steps as opposed to shrinking exponentially as in the usual tensor renormalization group (TRG). However, the cost of this new approach only scales as $O(\chi^4)$ in terms of the bond dimension $\chi$, significantly cheaper than the $O(\chi^6)$ cost scaling of TRG, whereas numerical benchmarking indicates that both approaches have comparable accuracy for the same bond dimension $\chi$. It follows that the new algorithm can allow for more accurate investigation of classical models on modestly sized lattices than is possible with standard TRG.
\end{abstract}

\maketitle

\section{Introduction}  \label{:sec Introduction}
In the past few decades tensor networks (TN) \cite{TN1,TN2,TN3} have become a popular tool in computational and theoretical physics, both for the study of classical and quantum many-body systems. Building on the tremendous success of matrix product states (MPS) \cite{MPS1,MPS2,MPS3} and the density matrix renormalization group (DMRG) \cite{DMRG1,DMRG2} algorithm in studying one dimensional systems, many other classes of tensor network ansatz and algorithms have been developed. Examples include tensor network ansatz such as the multi-scale entanglement renormalization ansatz (MERA) \cite{MERA,2DMERA1,2DMERA2,CritMERA} and the projected entangled pair states (PEPS)\cite{PEPS1,PEPS2,PEPS3}, as well as tensor network realizations of coarse-graining algorithms such as the tensor renormalization group (TRG) \cite{TRG,TRGplus,TERG, SRG,TEFR,TRGenv, TRG3D,SpinNet, TRG3DB,TRGcheap,TRGani} and tensor network renormalization (TNR) \cite{TNR1,TNR2,TNR3,TNR4} or similar methods \cite{TNRc1,TNRc2,TNRc3,TNRc4,TNRc5}. Despite ongoing efforts to improve computational performance, high computational cost is still the main obstacle for large scale calculations with many tensor network methods. Thus, even with the rapid development of computing power, there is a strong desire to reduce the computational cost through introduction of more efficient algorithms. 

In this work we introduce a novel coarse-graining algorithm, which we call \emph{core-tensor renormalization group} (CTRG), that can be applied to contract a $2D$ tensor network. Thus, similar the standard tensor renormalization group, this method could be used to study $2D$ classical many-body systems, where the network encodes the partition function, or $1D$ quantum systems, where the network encodes the Euclidean path integral given from a Suzuki-Trotter decomposition \cite{Suzuki1,Suzuki2}. However, the proposed CTRG method is demonstrated to be significantly more efficient than TRG in certain scenarios, allowing for more accurate numerical investigation of many-body systems, yet still retains the robustness and simplicity of implementation of the TRG approach.

The outline of our manuscript is as follows. We begin in Sect. \ref{:sec overview} with a non-technical overview of the proposed algorithm and its benefits, before providing the full implementation details in Sect. \ref{:sec CTRG steps}. Benchmark results and comparisons with TRG are provided in Sect. \ref{:sec benchmark results}, then discussions and conclusions are presented in Sect. \ref{:sec Discussion}.

\begin{figure}[!t!b]
	\centering
	\includegraphics[width=8.5cm]{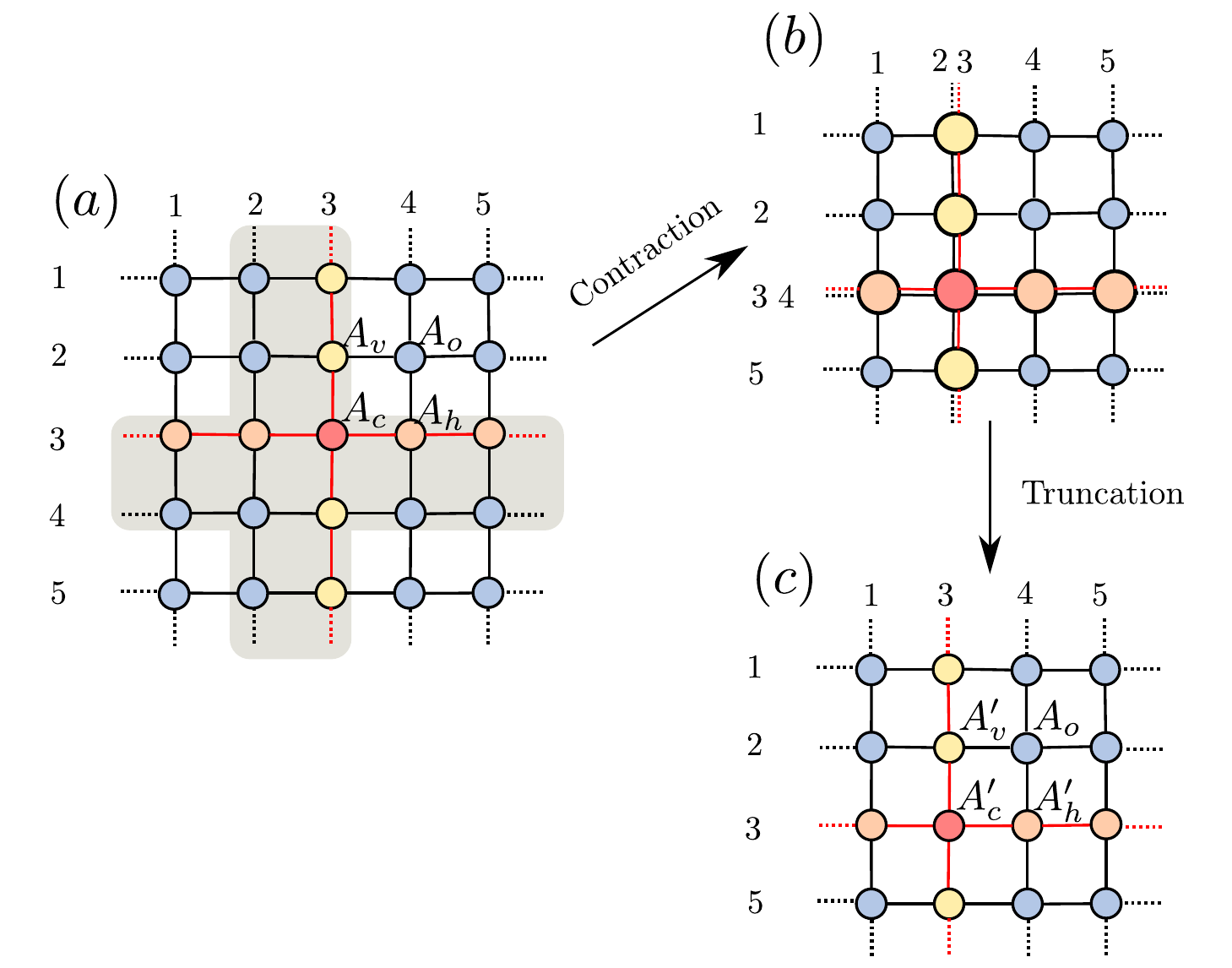}
	\caption{A depiction of the CTRG iteration, which maps an $L \times L$ lattice of tensors to an $(L-1) \times (L-1)$ lattice. (a) The initial network is everywhere composed of copies of the bulk tensor $A_0$, except for a single `core' row and column containing tensors $\{A_c, A_h, A_v \}$ as indicated. (b) An adjacent row and column of the network has been contracted into the core row/column, thus growing the index dimension of the core tensors. (c) The indices of the core tensors are truncated to dimension $\chi$, as to obtain new core tensors $\{A'_c, A'_h, A'_v \}$.}
	\label{:fig nontechnical}
\end{figure}

\section{Overview}\label{:sec overview}
In this section we provide an overview of the CTRG algorithm and discuss its cost scaling in comparison to TRG. As with TRG, the starting point of the CTRG algorithm is a $2D$ lattice of tensors. For simplicity we assume to have a homogeneous square lattice of linear dimension $L$ comprised of copies of a tensor $A_0$, assumed to have 4 indices of dimension $d$. Note that the method can also easily be extended to non-homogeneous networks as well as to other lattice geometries (e.g. hexagonal and triangular lattices). Through standard mappings, this network could encode either the partition of a $2D$ classical system on the square lattice, where the $A_0$ tensors encode the Boltzmann weights associated to plaquettes (see Appendix \ref{:appendix 2}), or encode the Euclidean path integral of a $1D$ quantum system, where the $A_0$ tensors encode the local gates given from a Suzuki-Trotter expansion \cite{Suzuki1,Suzuki2}. As such, contracting the $2D$ tensor network can be tantamount to calculating properties, such as expectation values, of $2D$ classical or $1D$ quantum many-body systems.

As a preliminary step of CTRG we select one tensor $A_0$ to act as the `core' tensor, which we re-label as $A_c$, and similarly re-label tensors in the same column as $A_v$ and tensors in the same row $A_h$. Then an iteration of the CTRG algorithm simply consists of the following two steps as depicted in Fig. \ref{:fig nontechnical}: (i) contract the adjacent row and column of the network into the core row/column, (ii) truncate the dimension of the core row/column, retaining at most $\chi$ dimensional indices, and obtaining new core tensors $\{A'_c, A'_v, A'_h\}$. Thus starting from an $L \times L$ lattice, which is homogeneous $A_0$ everywhere except for a single row and column, after a single step of CTRG we obtain an $(L-1) \times (L-1)$ lattice, which is again homogeneous $A_0$ everywhere except for a single row and column. It follows that after performing $L$ iterations of CTRG a network of $O(1)$ tensors is obtained, which can then be exactly contracted.

As will be shown in Sect. \ref{:sec CTRG steps}, the cost of a single iteration of CTRG scales as $O(\chi^4 d^3)$, with $d$ the index dimension of the initial network. By extension, the cost of contracting over a spatially homogeneous lattice of linear dimension $L$ using CTRG scales as $O( \chi^4 d^3 L)$. In comparison, the cost of TRG is known to scale as $O(\chi^6 \log L )$ for a homogeneous $L\times L$ lattice, with $\chi$ the TRG bond dimension. However, as will be demonstrated in the benchmark results of Sect. \ref{:sec benchmark results}, both CTRG and TRG give comparable accuracy when using the same bond dimension $\chi$. Thus one expects CTRG to be the more computationally efficient algorithm in the large bond dimension $\chi$ limit for systems where $L$ is not too large. Indeed, the benchmark results for the $2D$ classical Ising model at critical temperature show that, even for modestly large lattices of linear dimension $L=256$ Ising spins, the CTRG algorithm vastly outperforms standard TRG. In this setting the CTRG algorithm can reach a level of accuracy in \emph{less than a minute} of computation time that would require \emph{many hours} using standard TRG, and the disparity between computation times grows wider as the bond dimension $\chi$ is increased.

\begin{figure}
	\centering
	\includegraphics[width=8.5cm]{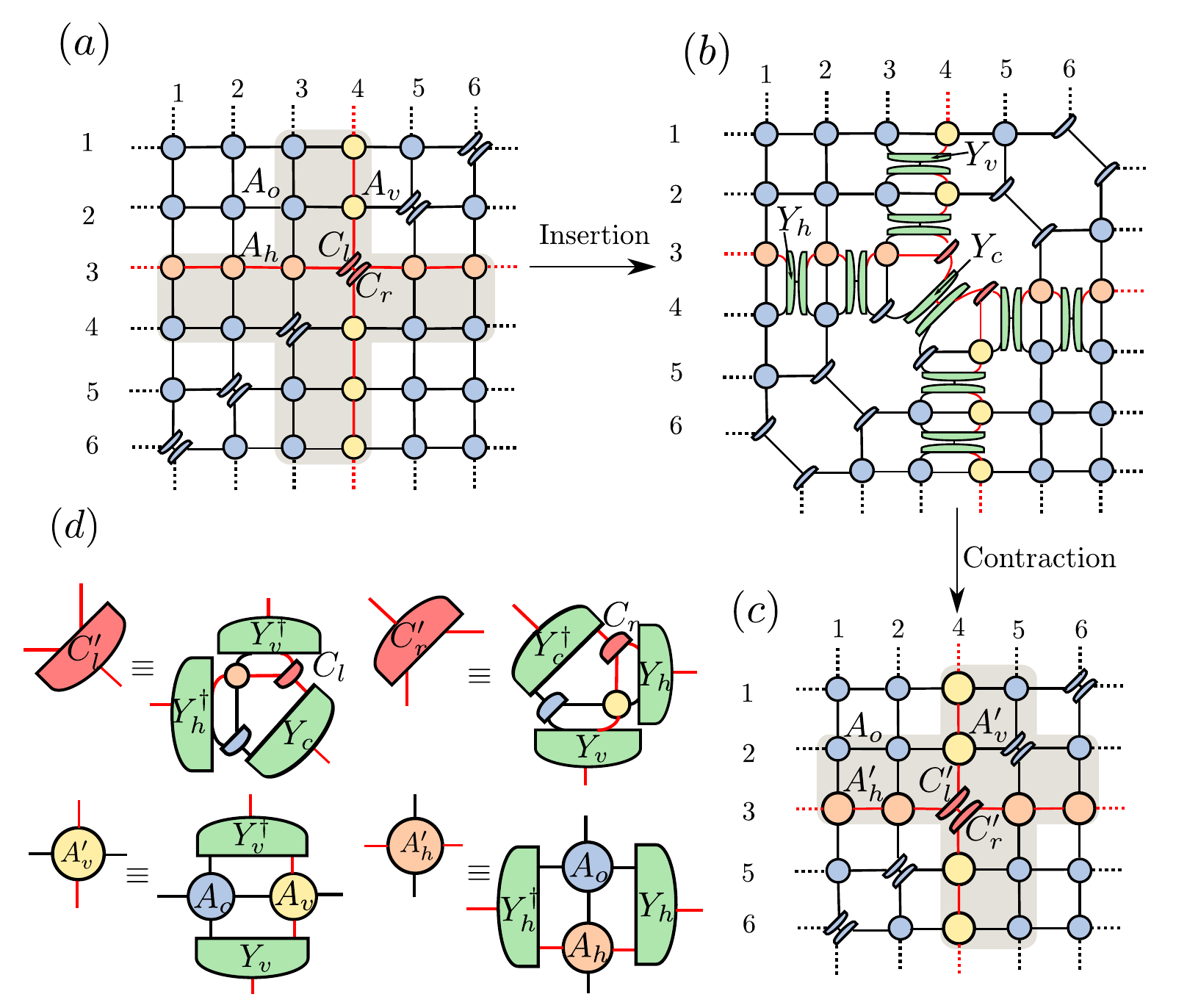} 
	\caption{At iteration of the CTRG algorithm. (a) The initial square lattice network is homogeneous except for a core row/column which contains core tensors $\{ A_v, A_h, C_l, C_r \}$ and a diagonal line through the core along in which the bulk tensors have been decomposed into products of 3-index tensors. (b) Pairs of isometries $\{ Y_v, Y_h, Y_c \}$ and their conjugates have been inserted into the core row/column of the network. (c) Isometries are contracted with their neighboring tensors, effectively absorbing a bulk row/column into the core row/column, as to produce new core tensors $\{ A'_v, A'_h, C'_l, C'_r \}$. (d) Definitions of the new core tensors.}
	\label{:fig rgsteps}
\end{figure}

\begin{figure}[h]
	\centering
	\includegraphics[width=8.5cm]{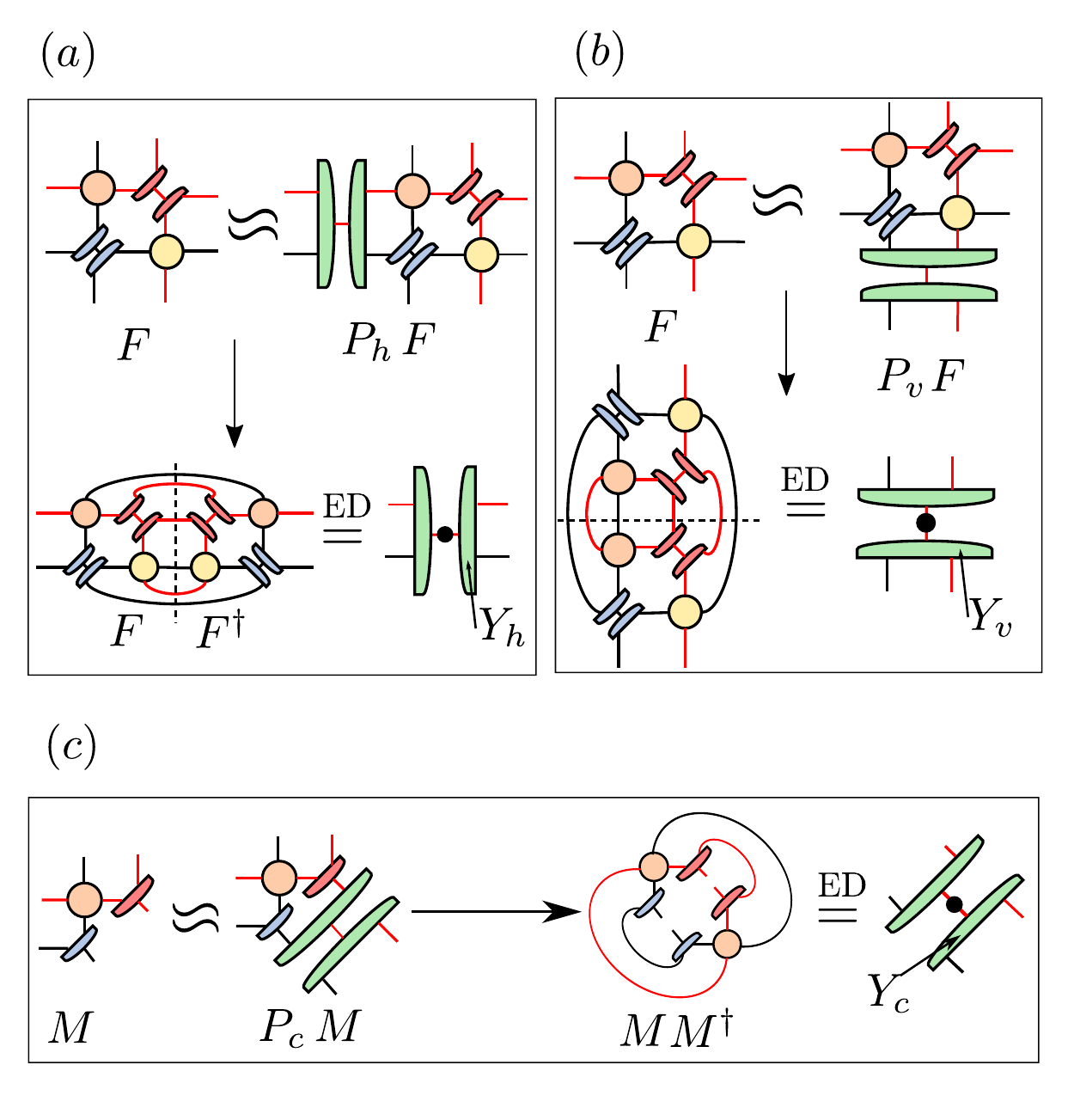}
	\caption{(a) The projector $P_h \equiv Y_h Y_h^\dag$ should be chosen to (approximately) leave invariant the network $F$, which is the network formed from the central tensors of the initial lattice in Fig. \ref{:fig rgsteps}(a). The optimal isometry $Y_h$ is formed by taking the eigenvalue decomposition (ED) of $F F^\dag$, when $F$ is viewed as a matrix between its left two and remaining indices, and truncating to retain only the $\chi$ dominant eigenvectors. (b) The optimal isometry $Y_v$ is obtained from the ED of $F F^\dag$, when $F$ is viewed as a matrix between its bottom two and remaining indices. (c) The optimal isometry $Y_c$ is obtained from the ED of $M M^\dag$, when $M$ is half of the $F$ network. }
	\label{:fig isometries}
\end{figure}

\section{Algorithm} \label{:sec CTRG steps}
In this section we describe in detail the proposed CTRG algorithm. As explained in Sect. \ref{:sec overview} the starting point of the algorithm is a square lattice tensor network composed of four index bulk tensors $A_0$ with index dimensions $d$. We select a single row and column to serve as the core, relabeling the column tensors $A_v$, the row tensors $A_h$, and the central tensor $A_c$. Here we also perform an additional step of decomposing $A_c$ into a pair of 3-index tensors $C_l$ and $C_r$ using the singular value decomposition (SVD), and similarly decompose certain bulk tensors $A_0$, located on the forward sloping diagonal through the core of the lattice, into pairs of 3-index tensors as indicated in Fig. \ref{:fig rgsteps}(a). As will be later evident, by representing the central tensor $A_c$ as the product of a pair of 3-index tensors $\{C_l,C_r\}$ the computational cost is reduced to $O(\chi^4)$, which would otherwise be $O(\chi^5)$ if instead $A_c$ was kept 4-index tensor. We now describe how a single iteration of CTRG, which maps an $L\times L$ lattice composed of tensors $\{A_0, A_v, A_h, C_l, C_r \}$ into an $(L-1)\times (L-1)$ lattice composed of tensors $\{A_0, A'_v, A'_h, C'_l, C'_r \}$, by appropriately absorbing an adjacent bulk row/column into the core row/column. For simplicity we describe only the step where the bulk row/column from the below/left of the core is absorbed, although in practice we alternate each iteration with absorbing the bulk row/column from the above/right of the core.

An iteration of the coarse-graining step is implemented using three different isometric tensors $\{Y_h, Y_v, Y_c\}$, whose output dimension is upper bounded at some specified maximal bond dimension $\chi$. The isometries $Y_h$ act to compress a bulk row into the core row, while isometries $Y_h$ act to compress a bulk column into the core column, and finally isometries $Y_c$ act to compress the index joining the central tensors $C_l$ and $C_r$, as depicted in Fig. \ref{:fig rgsteps}(b-c). Let us define the small network $F$ as that composed of the central tensors $\{C_l, C_r \}$ together with a single tensor $A_h$ and $A_v$ from the core column and row respectively, and a single bulk tensor $A_0$, as depicted in Fig. \ref{:fig isometries}. The isometries are chosen as to minimize the truncation error resulting from their insertion into the network; more specifically, they should leave network $F$ invariant when applied in conjugate pairs. For instance, if we define $P_h \equiv Y_h Y_h^\dag$ as the rank-$\chi$ projector defined from isometry $Y_h$ and its conjugate, then the truncation error $\varepsilon$ when acting on $F$ is given as,
\begin{equation}
\varepsilon  = \left\| {F - {P_h}F} \right\|, \label{eq:eps}
\end{equation}
where network $F$ has been considered as a matrix between its left two indices and the remaining indices, as depicted in Fig. \ref{:fig isometries}(a). Fortunately the solution for the optimal rank-$\chi$ projector $P_h \equiv Y_h Y_h^\dag$ that minimizes the error $\varepsilon$ in Eq. \ref{eq:eps} is easily obtained: isometry $Y_h $ should be formed from the $\chi$ dominant eigenvectors (i.e. those with largest magnitude eigenvalue) of the Hermitian matrix $F F^\dag$. Similarly, the optimal isometry $Y_v$ to minimize the truncation error is obtained through eigenvalue decomposition (ED) of $F F^\dag$ when network $F$ is treated as a matrix between its bottom two indices and the remaining indices, see Fig. \ref{:fig isometries}(b). Finally, the remaining isometry $Y_c$ is chosen to minimize the truncation error associated to the diagonal index joining the core tensors $C_l$ and $C_r$. This is accomplished by requiring the projector $P_c \equiv Y_c Y_c^\dag$ to minimize the error when acting on half of network $F$, split through the diagonal as depicted in Fig. \ref{:fig isometries}(c), again found through the appropriate eigenvalue decomposition.

Once the set of isometries $\{Y_h, Y_v, Y_c\}$ required for the coarse-graining step have been obtained they are then inserted into the network, as depicted in Fig. \ref{:fig rgsteps}(b), and contracted to form the coarse-grained core tensors $\{A'_v, A'_h, C'_l, C'_r \}$ as depicted in Fig. \ref{:fig rgsteps}(c-d). Thus the iteration is complete, and we can begin the next iteration to absorb another bulk row/column from the network into the core row/column.

Notice that during the coarse-graining step the same isometry $Y_h$ is used at every location within the row under consideration, likewise the same isometry $Y_v$ is used at every location within the column under consideration. This is despite the fact that the CTRG method appears to break the translation invariance of the initial network, where each tensor $A_h$ within the core row occupies a different location with respect to the intersection tensors $\{C_l, C_r \}$. This could lead one to believe (erroneously) that isometries $Y_h$ should vary with their relative position in the row. However our proposed approach is indeed justified since the core column after $z$ coarse-graining iterations is identical, up to truncation errors, to a width $(z+1)$ strip of tensors from the original lattice. Thus, up to small errors, tensors $A_h$ in the core row cannot `sense' the presence of the core column; it follows that all $A_h$ tensors possess an equivalent tensor environment regardless of their proximity to the center.

The key steps of the CTRG algorithm include (i) the contraction of the $F F^\dag$ networks as depicted in Fig. \ref{:fig isometries}, (ii) the eigenvalue decomposition of the $F F^\dag$ networks, and (iii) the contraction for new core tensors $\{A'_v, A'_h, C'_l, C'_r \}$ as depicted in Fig. \ref{:fig rgsteps}(d). Each of these steps can be accomplished with leading order computational cost $O(\chi^4 d^3)$ or less, with $d$ as the dimension of the original lattice and assuming that the indices of the core row/column have initial and final dimension $\chi$. In comparison, the computational cost of traditional TRG \cite{TRG} is known to scale as $O(\chi^6)$. The reduction in the cost as a power of $\chi$ when using CTRG can be attributed to the fact that only a single row/column of the bulk network is absorbed into the core at each step: the isometries only act to combine a $d$-dimensional and a $\chi$-dimensional index into a new $\chi$-dimensional index at each step. In contrast, the tensors used in TRG must act to combine two $\chi$-dimensional indices at each step, which is inherently more expensive as a function of $\chi$. 

\section{Benchmark results} \label{:sec benchmark results}
We test the efficiency of the proposed approach by benchmarking with the square-lattice classical Ising model, whose partition function is defined, 
\begin{align}
Z = \sum_{\{s\}} \left( \prod_{\bk{i,j}} \exp(-\beta s_is_j) \right), \;\; s_i \in \{+1,-1\} 
\end{align}
where $\bk{i,j}$ represents nearest-neighbor sites and $\beta$ is the inverse temperature, $\beta = 1/T$. We encode this partition function as a square-lattice tensor network using standard methodology \cite{TNR4}, see also Appendix \ref{:appendix 2}.

\begin{figure}[!t!b]
	\centering
	\includegraphics[width=8cm]{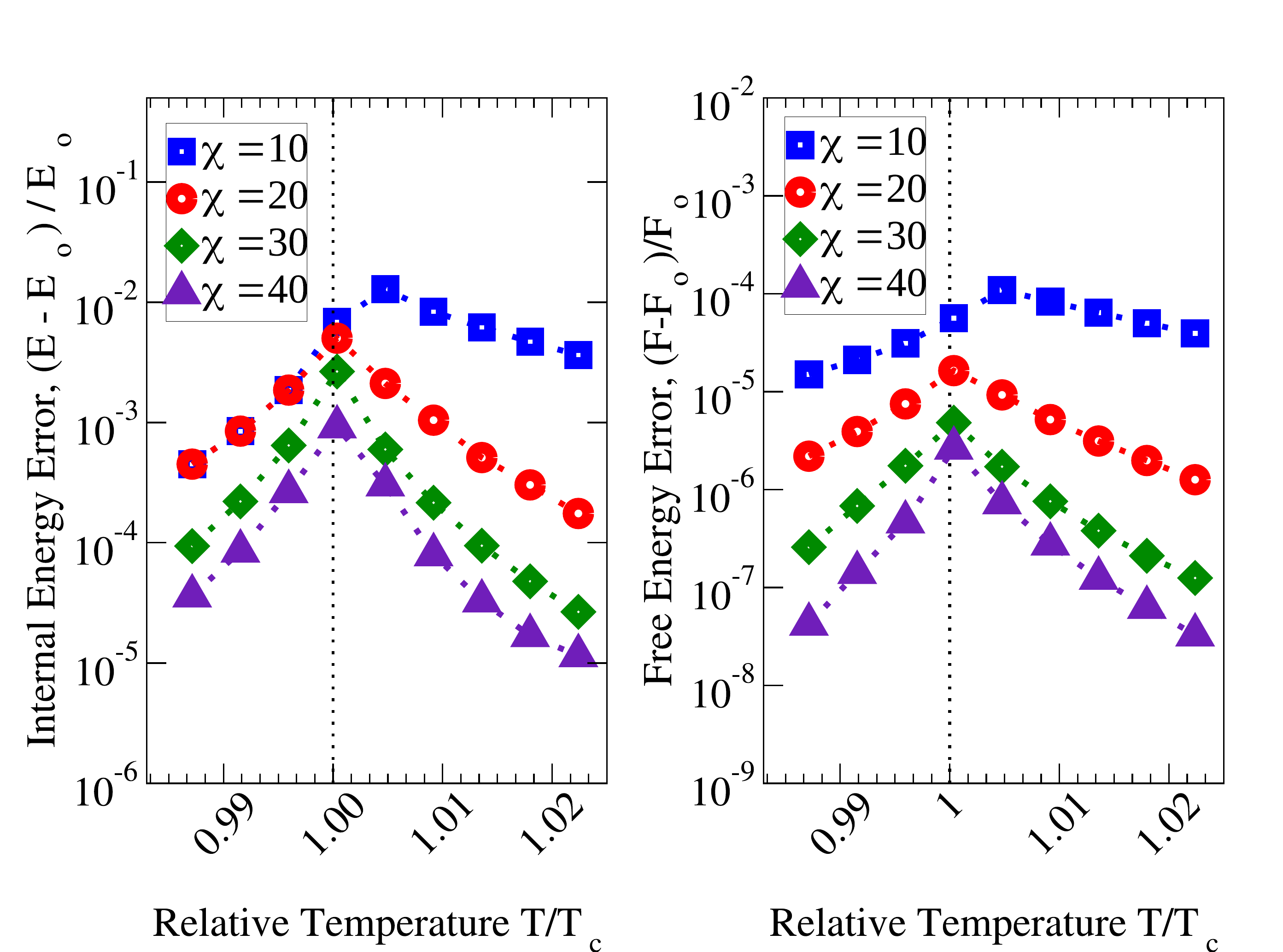}
	\caption{(a) Relative error in the (per-site) internal energy produced from CTRG applied to the Ising model in the thermodynamic limit, as a function of temperature $T$ relative to the critical temperature $T_c$ and of the bond dimension $\chi$. The errors are seen to be largest when close to $T_c$, but are systematically reduced as $\chi$ is increased. (b) The same calculations for the free energy density of the Ising model.}
	\label{:fig results}
\end{figure}

As a first test, we explore the accuracy of CTRG in terms of the free energy density and internal energy density, both as a function of temperature $T$ and of bond dimension $\chi$. In each calculation, we repeat the RG iteration until the energy densities are converged (i.e. such that the finite size effects are smaller than the accuracy threshold of the method) and then compare with the exact known Ising model energy densities in the thermodynamic limit. This required at most $3000$ RG steps to achieve convergence near the critical temperature, $T_c = 2/{\log{(1+\sqrt{2}})}$, which equates to lattices of linear size $L = 3000$ Ising spins. The results, as presented in Fig. \ref{:fig results}, are consistent with behavior of other tensor RG schemes \cite{TRG3D}: the accuracy is lowest when near $T_c$, but improves systematically everywhere in the phase diagram as the bond dimension $\chi$ is increased. 

\begin{figure}[!t!b]
	\centering
	\includegraphics[width=8cm]{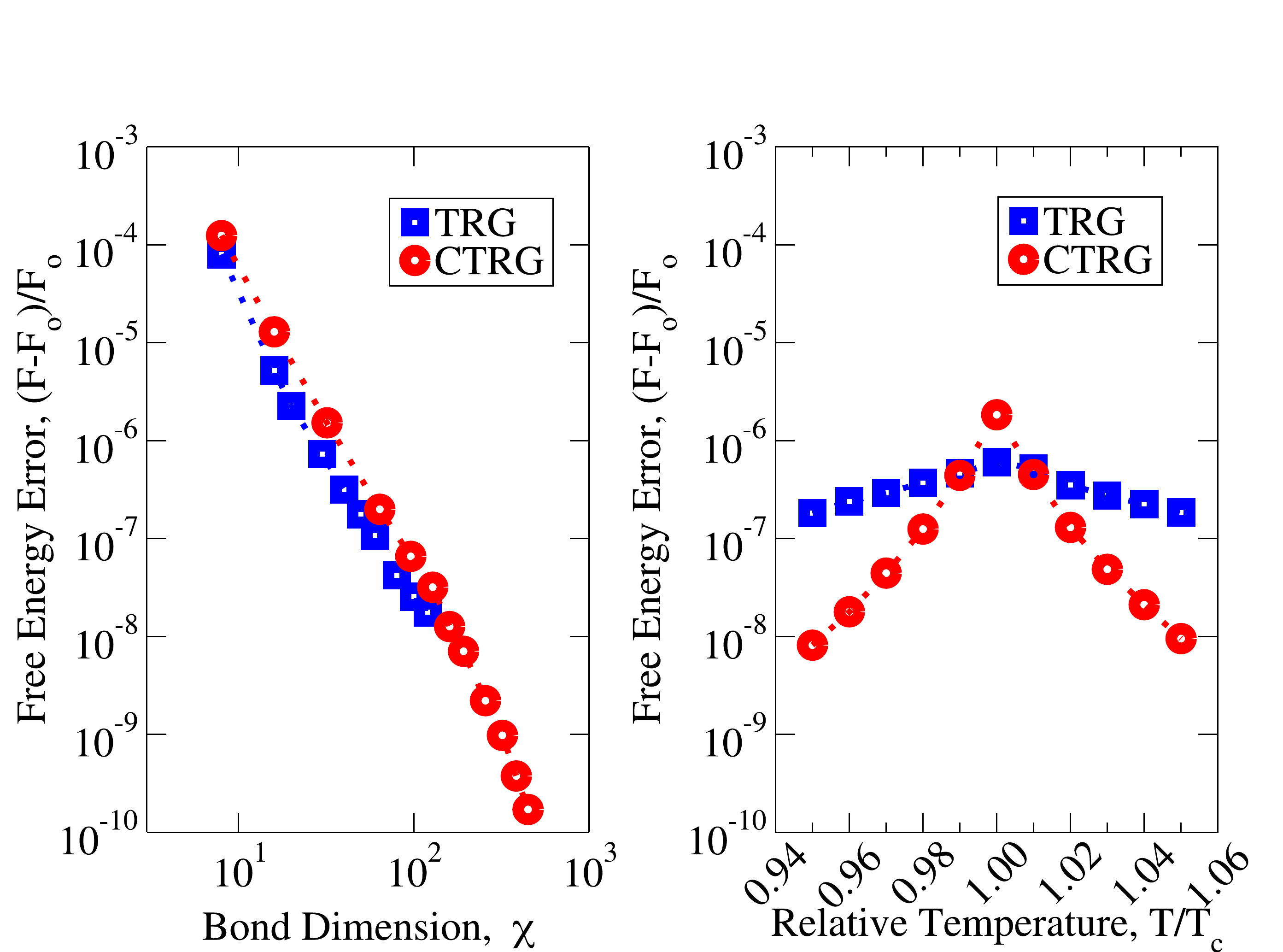}
	\caption{(a) A comparison of the accuracy of the free energy density produced by TRG and CTRG for the Ising model on an infinite strip of width $L=128$ sites at critical temperature. Both methods produce comparable accuracy for the same bond dimension $\chi$, with TRG giving only slightly more accurate energies. (b) Comparison between TRG and CTRG for accuracy of the free energy density as a function of temperature with fixed bond dimension $\chi=30$. }
	\label{:fig TRGvsNewRG}
\end{figure}

Next, in Fig. \ref{:fig TRGvsNewRG}, we compare the accuracy of the free energy given from CTRG against that given from a standard implementation of TRG \cite{TRG} for the Ising model on an infinite strip of width $L=128$. When using either method, the strip is mapped to an width $L=1$ chain of infinite length after a finite number of RG steps, which can then be solved using standard transfer matrix techniques. In lieu of an exact value for the free energy density on this strip, we use the free energy given from a $\chi = 640$ CTRG calculation as a proxy for the exact value (noting that the values plotted in Fig. \ref{:fig TRGvsNewRG} are converged with respect this reference energy). In Fig. \ref{:fig TRGvsNewRG} (a) we see that, for any fixed $\chi$, both TRG and CTRG yield very similar accuracies for the free energy at critical temperature $T_c$, with TRG only slightly more accurate for all $\chi$. Notable is that both methods converge polynomially to the exact free energy as a function of $\chi$, with the error in the free energy density $\Delta F$ scaling approximately as $\Delta F \propto \chi^{-3.4}$. Fig. \ref{:fig TRGvsNewRG}(b) compares the accuracy of TRG and CTRG at fixed bond dimension $\chi=30$ as a function of temperature. Interestingly, it is seen that CTRG gives greatly improved accuracy over TRG when away from $T_c$, closely matching the performance of more sophisticated versions of TRG that make use of a larger tensor environment in order to improve accuracy \cite{SRG,TRGenv}. 

\begin{figure}[!t!b]
	\centering
	\includegraphics[width=8cm]{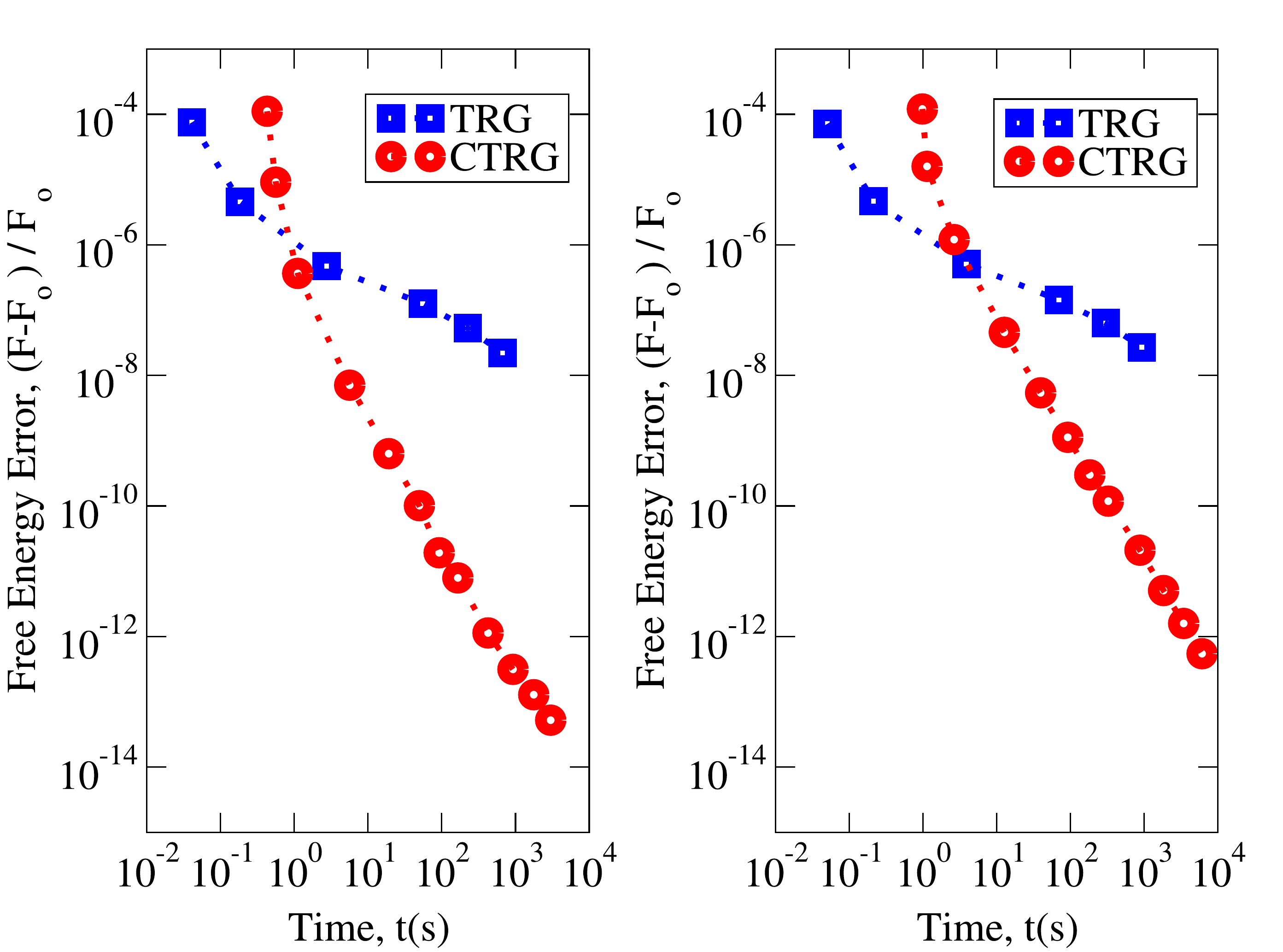}
	\caption{(a) Comparison between TRG and CTRG for accuracy in the free energy density as a function of computation time, for the Ising model at critical temperature on an infinite strip of width $L=128$ sites. For high-precision calculations CTRG is seen to be vastly superior in terms of computational efficiency. (b) Same comparison for the Ising model on an infinite strip of width $L=256$ sites.}
	\label{:fig time}
\end{figure}

Finally in Fig. \ref{:fig time} we compare the accuracy of TRG and CTRG as a function of computation time, with all calculations performed separately on the same hardware (3.4Ghz six-core desktop CPU), and for infinite length strips of width $L=128$ and $L=256$ Ising spins at critical temperature $T_c$. These results match the expectations that arise from consideration of the computational cost scaling of TRG, $O(\chi^6 \log L)$, versus that of CTRG, $O(\chi^4 L)$. Namely that for small $\chi$ calculations the $L$ component of the cost scaling dominates such that TRG is the more efficient approach, while in the large $\chi$ limit CTRG becomes the more efficient approach. However, we find it remarkable how early this crossover in efficiency occurs; on the width $L=256$ strip of Ising spins, CTRG is already the more efficient approach for calculations requiring more than a few seconds. Recently a version of TRG with a reduced computational cost scaling of $O(\chi^5 \log L)$ has been proposed in Ref. \onlinecite{TRGcheap}, which uses an iterative approach to find optimal projectors, rather than the SVD. Although this new variant of TRG would come closer to CTRG in terms of efficiency, it is still expected that it would be less efficient than CTRG in the large-$\chi$ limit. Note that all benchmark calculations were performed using code written in Julia (version 1.1.0) and made use of the `TensorOperations.jl' package.

\begin{figure}[!t!h!b]
	\centering
	\includegraphics[width=8cm]{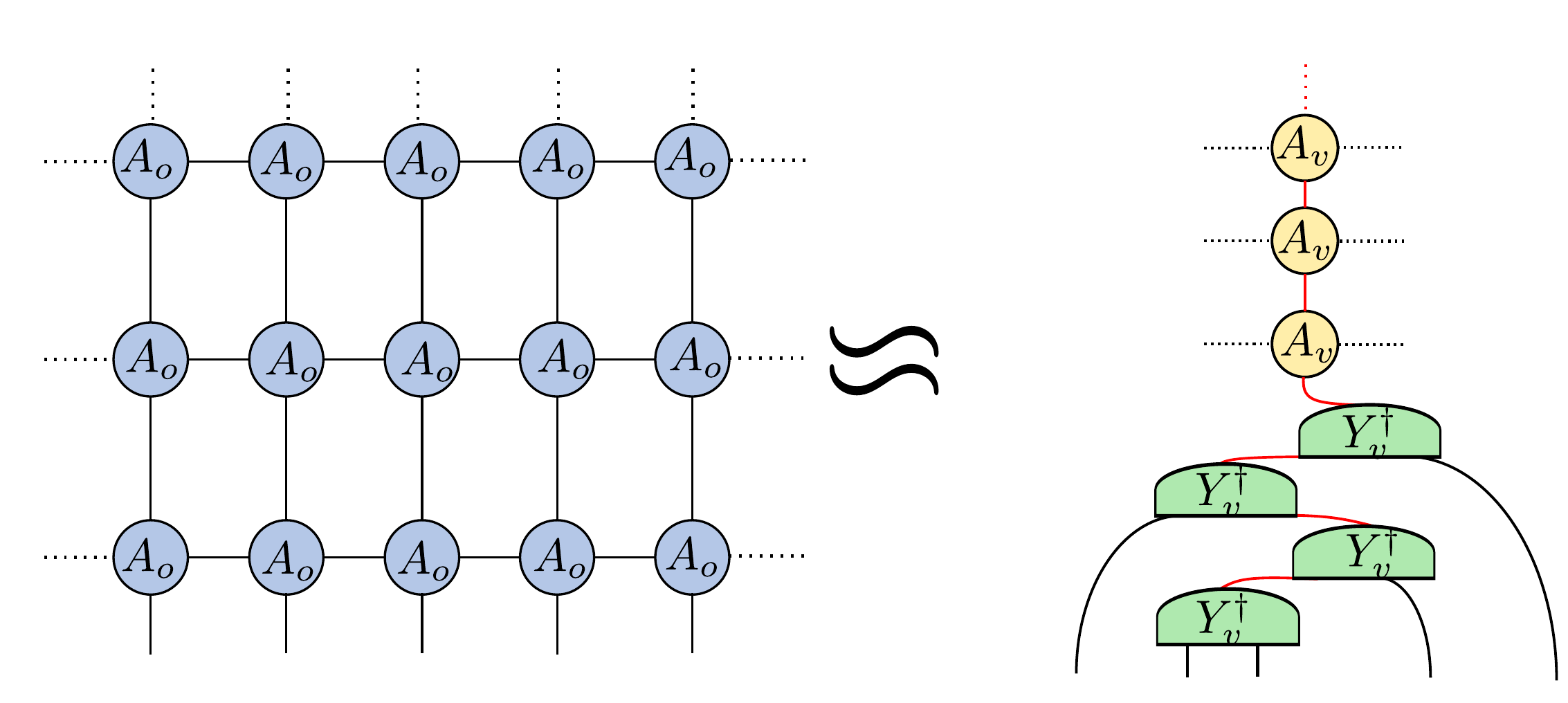}
	\caption{The CTRG algorithm can be understood as transforming the initial square-lattice network into a tree tensor network composed of isometries $Y_v$.}
	\label{:fig extension}
\end{figure}

\section{Discussion}\label{:sec Discussion}
An important observation that can be drawn from the benchmark results is that a CTRG calculation with bond dimension $\chi$ seems to perform comparably well to a TRG calculation of the same bond dimension, when applied to systems at or near criticality. This is not surprising since both methods are related to tree tensor network (TTN) ansatz \cite{TTN1,TTN2}. It can be understood that TRG produces an approximation to the initial square-lattice tensor network as a binary tree (an observation which is more readily apparent when considering the HOTRG \cite{TRG3D} version of the method). Similarly, it can be understood that CTRG produces an approximation to the initial width-$L$ lattice as a depth-$L$ tree, see Fig. \ref{:fig extension}, which would be equivalent to an MPS under folding through the center. Since both forms of tree can only support block entanglement entropy $S$ which is upper bounded as $S \le \log (\chi)$, it makes sense that they share the same accuracy limitations when applied to critical systems, which are known to have logarithmically diverging block entanglement entropy scaling \cite{Ent1,Ent2}.

A more recent advance in tensor RG schemes has been the introduction of a \emph{disentangling} step, as pioneered by tensor network renormalization (TNR) and related methods \cite{TNR1,TNR2, TNR3,TNR4,TNRc1,TNRc2, TNRc3,TNRc4,TNRc5}. TNR is known to produce an approximation to a square-lattice tensor network as a multi-scale entanglement renormalization ansatz (MERA) \cite{TNR2}, thus is a natural choice for investigating critical systems as it can support the correct entanglement scaling \cite{Ent1,Ent2}. A practical consequence of having the proper entanglement scaling is that the free energy converges exponentially in bond dimension $\chi$ to the exact value, i.e. as $\Delta F \propto \exp(-k \chi)$, as opposed to only polynomially quickly for TRG and CTRG. Thus, in the limit of large $\chi$, disentangling based methods such as TNR are still expected to outperform CTRG when applied to critical systems. However there are still some scenarios for which CTRG could the more efficient approach, such as (i) away from criticality where impact of disentangling is not so significant, or (ii) at intermediate values of $\chi$ where computational cost associated to disentangling outweighs the associated accuracy improvements. Moreover, from a practical standpoint, CTRG has the benefit of being much easier to implement than TNR and related approaches that use disentangling. This follows as CTRG inherits much of the simplicity of TRG; the isometries that implement each RG step are each given (deterministically) through a single eigen-decomposition as seen in Fig. \ref{:fig isometries}. In contrast the disentanglers required by TNR are given via an iterative method that repeats many times until the tensors are deemed sufficiently converged \cite{TNR4}. Such iterative methods have the possibility of becoming trapped in sub-optimal solutions, a possibility which is entirely avoided when using TRG or CTRG.

A potentially useful extension of CTRG would be towards networks in higher spatial dimension such as a cubic lattice tensor network, which could represent either the partition function of a $3D$ classical system or the path integral for a $2D$ quantum system. In this setting the HOTRG \cite{TRG3D} has proven to be a useful numerical tool, but it is hampered by a computational cost which scales as $O(\chi^{11})$ in terms of bond dimension $\chi$, although more efficient variations have recently been developed \cite{TRGani}. We hypothesize that a version of CTRG generalized for higher spatial dimensions could reproduce results of equivalent accuracy to HOTRG, but with a much lower cost scaling in bond dimension $\chi$. This possibility remains an intriguing avenue for future research. 

This research was undertaken thanks in part to funding from the Canada First Research Excellence Fund.

\appendix
\newpage

\section{Tensor Network Representation of Partition Functions} \label{:appendix 2}
\begin{figure}[h]
	\begin{center}
		\includegraphics[width=8cm]{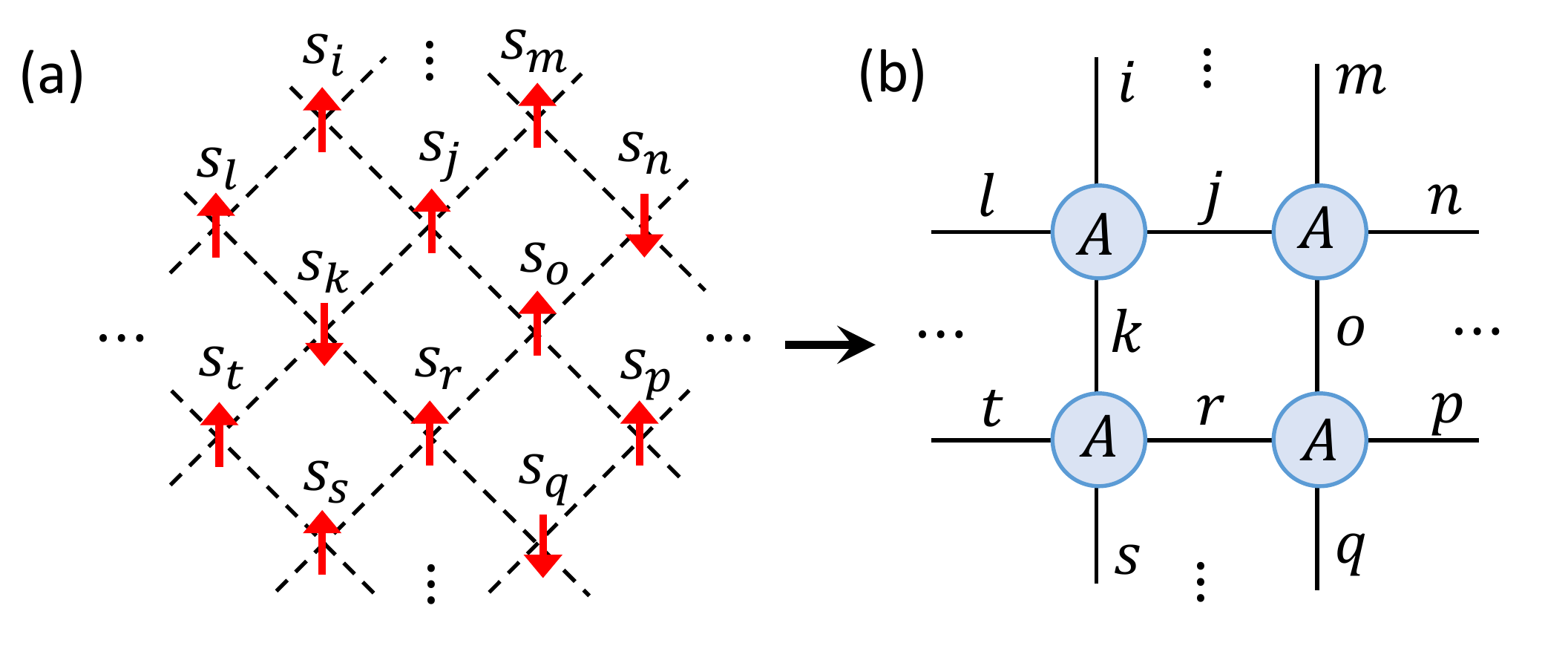}
	\end{center}
	\caption{(a) A square lattice of Ising spins, $s \in \{+1,-1\}$. (b) The partition function of the classical system can be encoded as a tensor network comprised of four-index tensors $A_{ijkl}$, where each tensor $A$ encodes the Boltzmann weights associated to the edges of the plaquette on which it is centered.} 
	\label{fig:Partition}
\end{figure}

In this appendix we describe how the partition function $Z$ of a $2D$ classical statistical system  at temperature $T$, 
\begin{equation} \label{eq:Partition}
Z = \sum_{\{\sigma\}} e^{-H(\{\sigma\})/T},
\end{equation}
can be expressed as a network of tensors. As a specific example we consider the square-lattice classical Ising model, with Hamiltonian functional,
\begin{equation} \label{eq:HamFunct}
H\left( \{ \sigma \} \right) =  - \sum_{\left\langle i,j \right\rangle}  \sigma _i\sigma _j,
\end{equation}
where $\sigma_i \in \{+1,-1\}$ is an Ising spin on site $i$. The partition function can be represented as a square-lattice tensor network composed of copies of a four-index tensor $A_{ijkl}$, where tensors sit in the center of every second plaquette of Ising spins according to a checkerboard tiling as depicted in Fig. \ref{fig:Partition}, such that there is one tensor $A$ for every two spins. Each tensor $A$ encodes the four Boltzmann weights $e^{\sigma_i\sigma_j/T}$ of the Ising spin interactions on the edges of the plaquette on which it sits,
\begin{equation} \label{eq:A}
A_{ijkl} = e^{\left(\sigma_i\sigma_j + \sigma_j\sigma_k + \sigma_k\sigma_l + \sigma_l\sigma_i \right)/T},
\end{equation}
such that the partition function $Z$ is thus as the sum over all indices from the network of $A$ tensors,
\begin{equation} \label{eq:Z}
Z = \sum_{ijk\cdots} A_{ijkl}  A_{mnoj}  A_{krst}  A_{opqr} \cdots. 
\end{equation} 
Notice that this method for constructing tensor networks representations of partition functions can be employed for any model with nearest-neighbor interactions. 

\begin{thebibliography}{99}
\bibitem{TN1}
J. I. Cirac and F. Verstraete, {\it Renormalization and tensor product states in spin chains and lattices}, J. Phys. A: Math. Theor. 42, 504004 (2009).
\bibitem{TN2}
R. Orus, {\it A practical introduction to tensor networks: Matrix product states and projected entangled pair states}, Ann. Phys. 349, 117 (2014).
\bibitem{TN3} 
J. C. Bridgeman and C. T. Chubb, {\it Hand-waving and Interpretive Dance: An Introductory Course on Tensor Networks}, J. Phys. A: Math. Theor. 50, 223001 (2017).

\bibitem{MPS1} 
M. Fannes, B. Nachtergaele, and R. F. Werner, {\it Finitely correlated states on quantum spin chains}, Commun. Math. Phys. 144, 443 (1992).
\bibitem{MPS2}
S. Ostlund and S. Rommer, {\it Thermodynamic limit of density matrix renormalization}, Phys. Rev. Lett. 75, 3537 (1995).
\bibitem{MPS3}
U. Schollw\"{o}ck, {\it The density-matrix renormalization group in the age of matrix product states}, Ann. Phys. 326, 96 (2011).

\bibitem{DMRG1}
S. R. White, {\it Density matrix formulation for quantum renormalization groups}, Phys. Rev. Lett. 69, 2863 (1992).
\bibitem{DMRG2}
S. R. White, {\it Density-matrix algorithms for quantum renormalization groups}, Phys. Rev. B 48, 10345 (1993). 

\bibitem{MERA}
G. Vidal, {\it A class of quantum many-body states that can be efficiently simulated}, Phys. Rev. Lett. 101, 110501 (2008).
\bibitem{2DMERA1}
L. Cincio, J. Dziarmaga, and M. M. Rams, {\it Multiscale entanglement renormalization ansatz in two dimensions: quantum Ising model}, Phys. Rev. Lett. 100, 240603 (2008).
\bibitem{2DMERA2} 
G. Evenbly and G. Vidal, {\it Entanglement renormalization in two spatial dimensions}, Phys. Rev. Lett. 102, 180406 (2009).
\bibitem{CritMERA} 
G. Evenbly and G. Vidal, {\it Quantum criticality with the multi-scale entanglement renormalization ansatz}, Chapter 4 in \textit{Strongly Correlated Systems: Numerical Methods}, edited by A. Avella and F. Mancini (Springer Series in Solid-State Sciences, Vol. 176 2013).

\bibitem{PEPS1} 
F. Verstraete and J. I. Cirac, {\it Renormalization algorithms for Quantum-Many Body Systems in two and higher dimensions}, arXiv:cond-mat/0407066 (2004).
\bibitem{PEPS2} 
F. Verstraete, J.I. Cirac, and V. Murg, {\it Matrix Product States, Projected Entangled Pair States, and variational renormalization group methods for quantum spin systems}, Adv. Phys. 57, 143 (2008).
\bibitem{PEPS3}
J. Jordan, R. Orus, G. Vidal, F. Verstraete, and J. I. Cirac, {\it Classical simulation of infinite-size quantum lattice systems in two spatial dimensions}, Phys. Rev. Lett. 101, 250602 (2008).

\bibitem{TRG} 
M. Levin and C. P. Nave, {\it Tensor renormalization group approach to 2D classical lattice models}, Phys. Rev. Lett. 99, 120601 (2007).
\bibitem{TRGplus}
H. C. Jiang, Z. Y. Weng, and T. Xiang, {\it Accurate determination of tensor network state of quantum lattice models in two dimensions}, Phys. Rev. Lett. 101, 090603 (2008).
\bibitem{TERG} Z.-C. Gu, M. Levin, and X.-G. Wen, {\it Tensor-entanglement renormalization group approach as a unified method for symmetry breaking and topological phase transitions}, Phys. Rev. B 78, 205116 (2008).
\bibitem{SRG} 
Z.-Y. Xie, H.-C. Jiang, Q.-N. Chen, Z.-Y. Weng, and T. Xiang, {\it Second Renormalization of Tensor-Network States}, Phys. Rev. Lett. 103, 160601 (2009).
\bibitem{TEFR} 
Z.-C. Gu and X.-G.Wen, {\it Tensor-Entanglement-Filtering Renormalization Approach and Symmetry Protected Topological Order}, Phys. Rev. B 80, 155131 (2009).
\bibitem{TRGenv} 
H.-H. Zhao, Z.-Y. Xie, Q.-N. Chen, Z.-C. Wei, J. W. Cai, and T. Xiang, {\it Renormalization of tensor-network states}, Phys. Rev. B 81, 174411 (2010).
\bibitem{TRG3D} 
Z.-Y. Xie, J. Chen, M. P. Qin, J. W.  Zhu, L. P. Yang, and T. Xiang, {\it Coarse-graining renormalization by higher-order singular value decomposition}, Phys. Rev. B 86, 045139 (2012).
\bibitem{SpinNet} 
B. Dittrich, F. C. Eckert, and M. Martin-Benito, {\it Coarse graining methods for spin net and spin foam models}, New J. Phys. 14 035008 (2012).
\bibitem{TRG3DB} 
A. Garcia-Saez and J. I. Latorre, {\it Renormalization group contraction of tensor networks in three dimensions}, Phys. Rev. B 87, 085130 (2013).
\bibitem{TRGcheap}
Y. Nakamura, H. Oba, and S. Takeda, {\it Tensor renormalization group algorithms with a projective truncation method}, Phys. Rev. B 99, 155101 (2019).
\bibitem{TRGani} 
D. Adachi, T. Okubo, and S. Todo, {\it Anisotropic Tensor Renormalization Group}, arXiv:1906.02007 (2019).

\bibitem{TNR1} 
G. Evenbly and G. Vidal, {\it Tensor network renormalization}, Phys. Rev. Lett. 115, 180405 (2015).
\bibitem{TNR2} 
G. Evenbly and G. Vidal, {\it Tensor network renormalization yields the multi-scale entanglement renormalization ansatz}, Phys. Rev. Lett. 115, 200401 (2015).
\bibitem{TNR3} 
G. Evenbly and G. Vidal, {\it Local scale transformations on the lattice with tensor network renormalization}, Phys. Rev. Lett. 116, 040401 (2016).
\bibitem{TNR4} 
G. Evenbly, {\it Algorithms for tensor network renormalization}, Phys. Rev. B 95, 045117 (2017).
\bibitem{TNRc1}
S. Yang, Z.-C. Gu, and X.-G. Wen, {\it Loop optimization for tensor network renormalization}, Phys. Rev. Lett. 118, 110504 (2017).
\bibitem{TNRc2}
M. Bal, M. Mari\"en, J. Haegeman, F. Verstraete, {\it Renormalization group flows of Hamiltonians using tensor networks}, Phys. Rev. Lett. 118, 250602 (2017).
\bibitem{TNRc3}
L. Ying, {\it Tensor Network Skeletonization}, Multiscale Model. Sim. 15-4 pp. 1423-1447 (2017).
\bibitem{TNRc4}
M. Hauru, C. Delcamp and S. Mizera, {\it Renormalization of tensor networks using graph independent local truncations}, Phys. Rev. B 97, 045111 (2018).
\bibitem{TNRc5}
G. Evenbly, {\it Gauge fixing, canonical forms, and optimal truncations in tensor networks with closed loops}, Phys. Rev. B 98, 085155 (2018).

\bibitem{Suzuki1}
M. Suzuki, {\it Fractal decomposition of exponential operators with applications to many-body theories and Monte Carlo simulations}, Phys. Lett. A, 146, 6, (1990) 319-323. 
\bibitem{Suzuki2}
M. Suzuki, {\it General theory of fractal path integrals with applications to many‐body theories and statistical physics}, J. Math. Phys. 32, 2, (1991) 400-407.

\bibitem{TTN1}
Y. Shi, L. Duan and G. Vidal, {\it Classical simulation of quantum many-body systems with a tree tensor network}, Phys. Rev. A 74, 022320 (2006).
\bibitem{TTN2}
L. Tagliacozzo, G. Evenbly and G. Vidal, {\it Simulation of two-dimensional quantum systems using a tree tensor network that exploits the entropic area law}, Phys. Rev. B 80, 235127 (2009).

\bibitem{Ent1} 
G. Vidal, J. I. Latorre, E. Rico, and A. Kitaev, {\it Entanglement in quantum critical phenomena}, Phys. Rev. Lett. 90, 227902 (2003).
\bibitem{Ent2}
J. I. Latorre, E. Rico, and G. Vidal, {\it Ground state entanglement in quantum spin chains}, Quant. Inf. Comput. 4, 48-92 (2004).

\end{thebibliography}
\end{document}